\pdfoutput=1
\documentclass[11pt,a4paper]{article}
\usepackage{amsmath,amssymb}
\usepackage{braket}
\usepackage{booktabs}
\usepackage{url}
\usepackage{mathtools}
\usepackage{amsmath}              
\usepackage{graphics,graphicx,epsfig,ulem} 
\usepackage{subfigure,subfig}
\usepackage{feynmp}
\usepackage{slashed}

\DeclareMathAlphabet{\mathbold}{U}{zeur}{b}{n}

\renewcommand{\Re}{\mathrm{Re}}

\renewcommand\[{\left[}
\renewcommand\]{\right]}

\def\beq{\begin{equation}}
\def\eeq{\end{equation}}
\def\[{\begin{equation}}
\def\]{\end{equation}}

\newcommand{\startappendix}{
\setcounter{section}{0}
\renewcommand{\thesection}{\Alph{section}}
\renewcommand{\theequation}{\Alph{section}.\arabic{equation}}}

\begin{document}
\numberwithin{equation}{section}

\title{
{\normalsize  \mbox{}\hfill IPPP/16/06, DCPT/16/12}\\
\vspace{2.5cm}
\Large{\textbf{Photo-production of a 750 GeV di-photon resonance \\ mediated by Kaluza-Klein leptons in the loop}}}

\author{Steven Abel and Valentin V. Khoze \\[4ex]
  \small{\it Institute for Particle Physics Phenomenology} \\
  \small{\it Durham University, Durham DH1 3LE, United Kingdom}\\[0.2ex]
  \small{\tt s.a.abel@durham.ac.uk, valya.khoze@durham.ac.uk}\\[0.8ex]
}

\date{}
\maketitle

\begin{abstract}
 \noindent We consider the phenomenology of a 750 GeV resonance $X$ which can
 be produced at the LHC by only photon fusion and subsequently decay into di-photons.
 We propose that the spin-zero state $X$ is coupled to a
 heavy lepton that lives in the bulk of a higher-dimensional theory and interacts only with
 the photons of the Standard Model. We compute the di-photon rate in these models with two and more
 compact extra dimensions and demonstrate that they allow for a compelling explanation of the di-photon excess
 recently observed by the ATLAS and CMS collaborations. The central role in our approach is played 
by the summation over the Kaluza-Klein modes of the new leptons, thus providing a significant enhancement of the
$X\to \gamma\gamma$ loops for the production and decay subprocesses. It is expected that the jet activity
accompanying these purely electromagnetic (at the partonic level) processes is numerically suppressed by factors 
such as $\alpha_{\rm em}^2 \,{\cal C}_{q\bar{q}}/{\cal C}_{\gamma\gamma}\sim 10^{-3}$.
 
 \end{abstract}

\bigskip
\thispagestyle{empty}
\setcounter{page}{0}

\newpage


\section{Introduction}
\label{sec:intro}

In the analysis of  the first data obtained in proton-proton collisions with  centre of mass energy $\sqrt{s}=13$ TeV at Run 2 of the LHC,
the ATLAS and CMS collaborations both reported an excess of events in the invariant mass distribution of two 
photons near 750 GeV \cite{ATLASdiphoton,CMSdiphoton}.

\medskip

If  confirmed by the bulk of the Run 2 LHC data in the near future, this excess indicates the existence of a new boson $X$
with mass around 750 GeV, that decays into two photons. Assuming that this is indeed the case, and not a statistical fluctuation,
we may finally have the first hint of physics beyond the Standard Model. 
This tantalising possibility led to an unprecedented explosion in the number of exploratory papers
dedicated to the 750 GeV di-photon resonance.\footnote{More than 150 di-photon papers have appeared in the first 4 weeks 
since December 15. We apologise in advance for citing only the papers which have directly influenced our work, and for any omissions.}

\medskip

The ATLAS collaboration  analysed 3.2 1/fb of data and reported an excess of 14 events 
at the di-photon invariant mass 
of 750 GeV, with a best-fit width of approximately 45 GeV. The quoted statistical significance of the ATLAS excess is
3.9 $\sigma$ or 2.3 $\sigma$ including the look-elsewhere effect.
The CMS collaboration reports an excess of 10 di-photon events with a narrow-width peak at around 760 GeV and a lower
statistical significance of  2.6 $\sigma$.
The cross-sections for the observed di-photon excess were estimated in Ref.~\cite{Franceschini:2015kwy} 
(see also Refs.~\cite{Ellis:2015oso,Gupta:2015zzs,Falkowski:2015swt,Altmannshofer:2015xfo,Davis:2016hlw}) as
\begin{eqnarray}
\sigma^{\rm ATLAS}_{pp \to \gamma\gamma} \,(13 {\rm TeV}) &=&
(10 \pm 3)\, {\rm fb}\,,
\label{eq:xcATLAS}\\
\sigma^{\rm CMS}_{pp \to \gamma\gamma} \,(13 {\rm TeV}) &=&
(6 \pm 3)\, {\rm fb}\,,\label{eq:xcCMS}
\end{eqnarray}
which we will take to be 8 fb. The di-photon resonance $X$ must be a boson thanks to the 
Landau-Yang theorem. In this work we will neglect the possibility of a spin-two particle
and focus on a spin-zero state -- a scalar or a pseudo-scalar.
As in most of the di-photon literature we will also assume the mass and the width of the $X$ to fit the ATLAS data, thus in total,
\beq
\sigma^{\rm obs}_{pp \to X\to \gamma\gamma} \,=\, 8\, {\rm fb}\,, \qquad
   M_{X}\,=\, 750\, {\rm GeV}\,, \qquad \Gamma_{\rm tot}\,=\, 45 \, {\rm GeV}\,,
\eeq
though the observational evidence for the latter feature of a relatively large total width $\Gamma_{\rm tot}/M_{X} \simeq\, 6\%$, is 
not strong; consequently in most of our numerical estimates in the next section
we will indicate an appropriate scaling factor between the 45 GeV value and a generic $\Gamma_{\rm tot}$.

\medskip

At first sight $X$ appears to be very similar to the Standard Model Higgs -- it is produced in $pp$ collisions
and its decay modes include the $\gamma\gamma$ channel. Hence it is natural to expect that it is dominantly produced from
initial protons via the gluon-gluon fusion process (plus contributions from quark-anti-quark annihilation) and then decays into 
two photons. 
However, because it is six times heavier than the Standard Model (SM) Higgs, the new boson $X$ provides an interesting complication to 
model-building -- it requires the addition of new mediators between $X$ and the SM, contributing to {\it both} the production 
and the decay process. One cannot hope to utilise any of the SM fermions either in the loop of the gluon fusion production of $X$, 
or in the $X \to \gamma \gamma$ decay loop. Since $M_X$ is above all of the SM fermion-anti-fermion thresholds,
$X$ bosons produced on-shell would rapidly decay into fermion-anti-fermion pairs at tree level, thus wiping out the corresponding
$gg \to X$ and $X\to \gamma \gamma$ branching ratios which are loop suppressed. 
Hence, both parts of the assumed $gg \to X\to \gamma\gamma$ process, the gluon fusion, and the di-photon decay have to be 
generated entirely by new heavy mediators with $M_{\rm med} > M_{X}/2$ propagating in the loops. 
Futhermore, the mixing of a scalar $X$ singlet with the SM Higgs must be severely suppressed
\cite{Falkowski:2015swt,Berthier:2015vbb},
$\sin^2 \theta_{\rm mix} < 10^{-2}$.
In summary, the entire parton-level process must be generated by Beyond SM (BSM) physics with no input from the Standard Model beyond just
providing the external states. In addition, the Yukawa couplings of $X$ to the mediators that are required to accommodate the observed di-photon 
rate for a 45 GeV resonance are already pushed to large values and become non-perturbative not much above the TeV scale 
\cite{Gu:2015lxj,Son:2015vfl,Goertz:2015nkp}.

\bigskip

In this paper we follow an alternative more minimal approach where the $X$ (pseudo)-scalar
communicates only to the $U(1)_Y$ hypercharge factor of the Standard Model. In this case one does not require
separate BSM-enabled production and decay mechanisms.
This idea that the di-photon resonance couples to photons but not gluons was put forward 
in two pioneering papers ~\cite{Fichet:2015vvy,Csaki:2015vek} already on December 17 2015, and considered further in 
Refs.~\cite{Anchordoqui:2015jxc,Csaki:2016raa,Fichet:2016pvq}.

The main theoretical challenge facing this `pure photon' set-up is that the di-photon channel is subdominant, 
being suppressed by $\alpha_{\rm em}^2$ relative to the gluon fusion rate -- and this,
as we will briefly review in Section {\bf \ref{sec:estimates}},
puts the model in dangerous territory, with a combination of non-perturbative couplings, very large numbers $N_f$ of mediator flavours 
and very low scales of new physics.

We will propose that these short-comings can be resolved by allowing a lepton mediator that feels
large extra dimensions. The resulting tower of Kaluza-Klein (KK) states quite easily enhances the di-photon rate sufficiently to match 
with observations. In Section {\bf \ref{sec:model}} we will examine the effect in various dimensions determining how the photon-fusion 
rates match with the expectations for the observed di-photon resonance. Following that, in section {\bf \ref{sec:P}} we will derive additional
constraints on the UV-cutoff of the theory by requiring perturbativity of the gauge coupling. Even though the assumptions about the content
of the theory are the most minimal, we find that a significant window remains to fit the di-photon signal in  perturbative settings.

The effect of radiative corrections on the scalar boson mass and its finite naturalness are discussed in
Section {\bf \ref{sec:MX}}, in a framework where a supersymmetric theory is compactified down to 4D with
Scherk-Schwarz boundary conditions. Following this, in Section~{\bf \ref{sec:dark}}, we
address the additional  invisible decays of $X$ to $\gamma\gamma$, that can account for the full total width 
$\Gamma_{\rm tot} = 45$ GeV, and hence enable the di-photon resonance to play a role as mediator between the SM 
and dark matter (DM) sectors. Finally Section~{\bf \ref{sec:conc}} presents our conclusions.


\section{From an EFT vertex to a vector-like lepton mediator}
\label{sec:estimates}

For a (pseudo)-scalar of mass $M_X$ and width $\Gamma_{\rm tot}$ one can express the di-photon cross-section 
using the standard narrow-width approximation expression,
\beq
\sigma_{pp \to X\to \gamma\gamma} \,(s)\,=\, \frac{1}{s\, M_X\, \Gamma_{\rm tot}}\left(\sum_{i=g,q,\gamma}
{\cal C}_{ii} \Gamma_{X\to ii}\right)\, \Gamma_{X\to \gamma\gamma}\,
\label{eq:narw}
\eeq
where the sum is over all partons  and ${\cal C}_{ii} $ are the dimensionless integrals over the corresponding parton
distribution functions. 
 For example for gluons ${\cal C}_{gg} $ at 13 TeV
was estimated in \cite{Franceschini:2015kwy} as
${\cal C}_{gg}\simeq 2137$, for quarks ${\cal C}_{u\bar{u}}\simeq 1054$, ${\cal C}_{d\bar{d}}\simeq 627$,  and for
photons ${\cal C}_{\gamma \gamma}\simeq 54$.

\medskip

We now concentrate on the case where the $X$ boson is coupled only to photons, thus keeping only
${\cal C}_{\gamma\gamma}$  in the sum on the right hand side of \eqref{eq:narw}.
In fact, one should also consider corrections to this process sourced by the photon pdf integral ${\cal C}_{\gamma\gamma}$,
by the `VBF' type processes, which are proportional to quark pdfs ${\cal C}_{q\bar{q}}$ with each of the
initial quarks emitting a photon. A very rough estimate for the relative importance of this effect would be
$({\cal C}_{q\bar{q}} /{\cal C}_{\gamma\gamma})\, \alpha_{\rm em}^2\,\sim 10^{-3}$, using the values quoted in \cite{Franceschini:2015kwy}.
While a detailed calculation of such effects is beyond the scope of this paper, we will take the above estimate
as a hint that the corrections to the leading photo-production process are relatively small and as a result the presence of additional
jets (in the VBF case arising from the initial quark partons) is suppressed relative to what one would expect 
for a heavy Higgs-like scalar production in the Standard Model.

Photon-photon fusion in elastic $pp$ scattering as searches of new physics
 was considered even earlier in  Ref.~\cite{Jaeckel:2012yz}.
But in the weakly-coupled SM-like settings, the branching ratio of a scalar resonance to photons is tiny, and hence the 
rate for such processes is negligible. For example, the SM Higgs production via photon fusion in elastic $pp$ collision is 
only $\sim 0.1$ fb \cite{Khoze:2001xm}.

The next step is to account for contributions from inelastic as well as elastic $pp$ collisions. 
Following Refs.~\cite{Fichet:2015vvy,Csaki:2016raa,HLKR} we will express the answer in the form
\beq
\sigma_{pp \,:\, \gamma\gamma \to X\to \gamma\gamma} \,(13 \, {\rm TeV})\,\simeq\,
10\, {\rm pb}\,\left( \frac{\Gamma_{\rm tot}}{45\, {\rm GeV}}\right)\, \left({\rm Br}_{X\to \gamma\gamma}\right)^2\,.
\label{eq:xc}
\eeq
The numerical factor $\sigma_0 = 10$ pb on the right hand side includes the contribution from the integral over the parton distribution functions
of the photons in the initial state protons. As explained in Refs.~\cite{Fichet:2015vvy,Csaki:2016raa,HLKR}, this factor can suffer from a
large theoretical uncertainty and arises from accounting for inelastic collisions where one or both initial
protons gets destroyed after emitting a photon. The estimates for $\sigma_0$ obtained in the recent literature give
$\sigma_0=10.8\, {\rm pb}^{\rm Ref. \cite{Csaki:2016raa}}$, $\sigma_0=8.2\, {\rm pb}^{\rm Ref. \cite{Fichet:2015vvy}}$, 
$\sigma_0=7.5\, {\rm pb}^{\rm Ref. \cite{Franceschini:2015kwy}}$ and the most recent calculation 
\cite{HLKR} gives $\sigma_0=4.1\, {\rm pb}$.
(An even earlier estimate made in \cite{Csaki:2015vek}, which was based on the (subdominant) elastic $pp$ collisions, gives the rate 
suppressed by two orders of magnitude relative to \eqref{eq:xc}.)
The overall factor of the order-10 pb is perhaps on the optimistic side, 
but for the purposes of our work which aims instead to enhance the ${\rm Br}_{X\to \gamma\gamma}$ fractions, it will not be critical if the 
overall coefficient in \eqref{eq:xc} is reduced.

The photon fusion cross section computed in the same manner at $8$ TeV in \cite{Fichet:2015vvy,Csaki:2016raa,Franceschini:2015kwy}
is roughly a factor of 2 smaller than the corresponding $13$ TeV 
result in \eqref{eq:xc}, which is consistent with the absence of the di-photon resonance signal in Run 1 LHC searches.

To match the signal rate quoted in \eqref{eq:xcATLAS}-\eqref{eq:xcCMS},
a rather large value of $X$ to photons branching fraction is required on the right hand side of \eqref{eq:xc},
\beq
{\rm Br}_{X\to \gamma\gamma}\,:=\, \frac{\Gamma_{ \gamma\gamma}}{\Gamma_{\rm tot}}\, =\,
(0.02 - 0.04)\, \sqrt{\frac{45 \, {\rm GeV}}{\Gamma_{\rm tot}}}\,,
\label{eq:bound}
\eeq
which in the case of the relatively wide 45 GeV resonance, as preferred by the current ATLAS data, 
and for the 8 fb signal rate amounts to 0.028 i.e. 2.8\%.
In the usual four-dimensional QFT settings, such large branching fractions to photons would require introducing new physics
already at the electro-weak scale.
To see this, we parametrise the interactions of $X$ with photons via the leading-order dimension-5 operators,
\beq
\frac{e^2 }{2\Lambda_{\rm sc}}\, X F^{\mu\nu} F_{\mu\nu}\,\, , \qquad
\frac{e^2}{2\Lambda_{\rm ps}}\, X F^{\mu\nu} \tilde{F}_{\mu\nu}\,,
\label{eq:EFT}
\eeq
 where $X$ is assumed to be a scalar in the first case, and a pseudo-scalar in the second case,
 and $\Lambda_{\rm sc/ps}$ is the new physics scale responsible for generating these EFT vertices.\footnote{As usual, such an
 EFT parametrisation would be meaningful only for $\Lambda$ greater than both, the $X$-resonance mass, $M=750$ GeV, and the
 electroweak scale $v$ itself.}
 For the partial width
 of $X$ to photons one has,
 \beq
 \Gamma_{ \gamma\gamma}\,=\, \pi \alpha_{\rm em}^2 \frac{M_X^3}{\Lambda^2}\,,
 \eeq
where $\Lambda$ is either of the two: $\Lambda_{\rm sc}$, $\Lambda_{\rm ps}$; and to achieve the required branching ratio in \eqref{eq:bound}, we need
\beq
\Lambda \,=\, \alpha_{\rm em} \,  \sqrt{\frac{\pi \cdot 750}{0.028 \cdot 45}} \, 
\left(\frac{45\, {\rm GeV}}{\Gamma_{\rm tot}}\right)^{1/4}\times 750\, {\rm GeV}\, \simeq\,  236\, {\rm GeV} \,,
\label{eq:estL}
\eeq
with the final expression assuming the 45 GeV resonance. Thus, we see that to reproduce the experimental signal (or excess)
attributed to the di-photon resonance with $M=750$ GeV and $\Gamma_{\rm tot}=45$ GeV in terms of the {\it photon fusion} process
on its own, would require the introduction of new physics degrees of freedom already at the electroweak scale, i.e. $\Lambda \sim v < M$, which is obviously at odds with the experiment.

If the total width of the di-photon resonance turns out to be much less than 45 GeV, the value of $\Lambda$ will be rescaled as indicated in
\eqref{eq:estL}. What is the minimal possible value of the total width? Clearly it is achieved when the branching ratio to photons
${\rm Br}_{X\to \gamma\gamma} \to 1$. It then follows from \eqref{eq:xc} and \eqref{eq:xcATLAS}, \eqref{eq:xcCMS} that
\beq
(\Gamma^{\rm min} _{\rm tot} / 45\, {\rm GeV} )\,\simeq\, 8 \cdot 10^{-4}\,.
\label{eq:minw}
\eeq

If we restore other theoretical and experimental uncertainties, 
such as the numerical factor $\sigma_0$ on the right hand side of \eqref{eq:xc}, and the total experimental cross-section
in \eqref{eq:xcATLAS}, \eqref{eq:xcCMS} which we denote $\sigma^{\rm obs}$, the estimate in \eqref{eq:estL}
becomes
\beq
\Lambda \,=\, 
\left(\frac{\sigma_0}{10\, {\rm pb}}\right)^{1/4} 
\left(\frac{8\, {\rm fb}}{\sigma^{\rm obs}}\right)^{1/4} 
\left(\frac{45\, {\rm GeV}}{\Gamma_{\rm tot}}\right)^{1/4} \times 236\, {\rm GeV} \,.
\label{eq:estL2}
\eeq
The quartic roots ensure that the deviations even by an order of magnitude 
in any of the three parameters away from the  `central' values would not lead to any significant deviations from the
bound by the electroweak scale in Eq.~\eqref{eq:xc}.\footnote{The only
exception is provided by an extremely narrow resonance where the width value can be reduced by up to 3 orders of magnitude.
In the extreme case of the minimal width \eqref{eq:estL}.
one could raise the the upper bound on $\Lambda$ to 1.4 TeV.}

\medskip

In fact, it is easy to see that the actual mass scale of the new physics degrees of freedom appearing in a hypothetical
four-dimensional perturbative 
extension of the SM appears to be further suppressed compared to the  estimate in \eqref{eq:estL} by
a multiplicative factor of $\sim 1/(12\pi^2)$, cf Eq.~\eqref{eq:MLest} below.

To verify this, consider augmenting the Standard Model by a massive Dirac fermion $L_D$ 
charged only under hypercharge $U(1)_Y$ and
coupled to the spin-zero SM-singlet $X$,
\begin{eqnarray}
{\cal L}_{X_{\rm sc}} &=& \bar{L}_D\left(i\gamma^\mu D_\mu -M_L\right) L_D
\,+\, y X \bar{L}_D L_D \,,
\label{eq:Xsc}\\
{\cal L}_{X_{\rm ps}} &=& \bar{L}_D\left(i\gamma^\mu D_\mu -M_L\right) L_D
\,+\, y X \bar{L}_D i\gamma^5L_D \,.
\label{eq:Xps}
\end{eqnarray}
Here $D_\mu= \partial_\mu -i g' Q B_\mu$ is the $U(1)_Y$ covariant derivative and $Q$ is the hypercharge of $L_D$, $M_L$ is its mass
and $y$ is the Yukawa coupling to the singlet $X$.
The equation \eqref{eq:Xsc} describes the scalar $X$ while \eqref{eq:Xps} corresponds to the pseudo-scalar di-photon resonance.

The Dirac fermion, $L_D$ consists of two 2-component Weyl spinors, $L_\alpha$ and $ \tilde{L}^{\dagger\, \dot{\alpha}}$
where $L$ transforms in fundamental, and 
$\tilde{L}$ in anti-fundamental representations, in this case only of the 
hypercharge, being a singlet under the $SU(3)_{\rm c}$ and the $SU(2)_{\rm L}$ gauge factors.

The $X\leftrightarrow \gamma\gamma$ process is described by the triangle diagram with the virtual $L_D$ fermion propagating in the loop.
Evaluating the diagram for the scalar $X$ using the interactions in \eqref{eq:Xsc}, one recovers the well-known result \cite{Spira:1995rr} for the 
$X FF$ formfactor, which we present as the expression for the Wilson coefficient of the first operator in \eqref{eq:EFT}:
\beq
\frac{e^2}{2\Lambda_{\rm sc}}\,=\,\, \frac{1}{M_L}\,\frac{e^2 Q^2 y}{24 \pi^2} \,\times {\cal F}(\tau) 
\label{eq:formfactor}
\,.
\eeq
Here,
\beq
\tau =\,\frac{M_X^2}{4M_L^2}\,\,, \quad {\rm and} \quad
{\cal F}(\tau) \,=\, \frac{3}{2 \tau^2} \left(\tau +(\tau-1) {\rm arcsin}^2(\sqrt{\tau})\right)\,.
\label{eq:formfactor2}
\eeq
The expression on the right hand side of \eqref{eq:formfactor} is presented in a factorised form, such that the formfactor 
${\cal F}(\tau)$ is normalised to one and tends to
$\to 1 + 7 \tau/30 +\ldots$ in the heavy lepton-mass limit $\tau \to 0$. Hence, in this limit we have a relation
\beq
M_L \,=\, N_f \, Q^2\, y\, \frac{\Lambda_{\rm sc} }{12 \pi^2}\,\, {\cal F}(\tau)\,,
\label{eq:MLest}
\eeq
where we have assumed that there are $N_f$ flavours of the vector-like leptons of mass $M_L$. 
Using the bound
$\Lambda\simeq 236$ GeV in
\eqref{eq:estL}, we are led to an estimate for an effective number of new lepton flavours,
\beq 
N_f \, \simeq\, 500\, \left(\frac{1}{Q^2}\right)\left(\frac{1}{y}\right)\left(\frac{M_{L}}{1\, {\rm TeV}}\right)\,,
\label{eq:MLest2}
\eeq
which - interpreted in a four-dimensional perturbative model with the TeV-scale vector-like leptons - is unjustifiably high.

\medskip

Such a large value for the number of  new lepton flavours to match
the observed di-photon excess \cite{ATLASdiphoton} by photon fusion is clearly unsatisfactory
and begs for an alternative explanation.
We will propose in the following section that the virtual new lepton states propagating in the
 loop extend into  large extra dimensions. This gives an infinite 
tower of Kaluza-Klein states of vector-like leptons populating the loops and, as we will show in the following section, 
leads to an appropriate enhancement necessary (and sufficient) to photo-produce
the observed di-photon rate. At the same time, the Kaluza-Klein lepton tower also contributes to the coupling constants of the theory 
which, not surprisingly, exhibit power-like (rather than merely logarithmic) growth at energies above the KK mass threshold. 
The theory tends to become strongly coupled at large UV scales. If one requires perturbativity of the theory, this amounts to
additional
constraints on the UV-cutoff of the theory $M_{|rm st} \lesssim 3$ TeV as will be shown in section {\bf \ref{sec:P}} 
based on the analysis of the gauge coupling. Even in this simplistic minimal theory, we find that a significant window remains 
to fit the di-photon signal in  perturbative settings.

\medskip

Before closing this section, we note that 
when the di-photon mediator $X$ is a pseudo-scalar, 
the calculation is essentially identical to the scalar case considered above.
A completely analogous result to \eqref{eq:formfactor} holds for the pseudo-scalar $X$ di-photon interaction following \eqref{eq:Xps},
\beq
\frac{e^2}{2\Lambda_{\rm ps}}\,=\,\, \frac{1}{M_L}\,\frac{e^2 Q^2 y}{16 \pi^2} \,\times {\cal F}_{\rm ps} (\tau)
\label{eq:ps_formfactor}
\,,
\eeq
only the functional form of the formfactor ${\cal F}_{\rm ps} (\tau)$ is different from the scalar case. But as we are
working in the limit of heavy lepton masses (the KK modes are becoming increasingly heavy), the formactor can again be set to one, and the same estimate \eqref{eq:MLest}-\eqref{eq:MLest2} applies.

\medskip

\section{A Simple Model with New Leptons in Large Extra Dimensions}
\label{sec:model}

As outlined in the Introduction, to address these shortcomings we wish to consider a KK tower of vector-like leptons coupled 
to photons and to $X$. In the simplest scenario, we take the new leptons to extend into the bulk of a $D$-dimensional theory 
\cite{Antoniadis:1990ew,ArkaniHamed:1998rs,Antoniadis:1998ig}
with $d=D-4$ flat compact extra dimensions.
We denote the D-dimensional coordinates,
\beq
 (x^0,x^1,x^2,x^3,z_1,\ldots z_d) \,\equiv\, (x^\mu,z_1,\ldots z_d) \,, \quad {\rm where} \quad
 0\,\le z_i \, \le \, 2\pi R_i \,.
 \eeq
The vector-like leptons are the only bulk fields we need to consider, and we assume them to be
charged only under the SM $U(1)_Y$ hypercharge, and to be colour- and $SU(2)_L$-singlets. Apart from these new leptons, we can 
take the entire Standard Model to be localised in four dimensions. However it is possible (and probably more natural in string models 
based on webs of D-branes) for the hypercharge to also be a bulk field. It will not make any difference to the discussion. 

\medskip

The fermions of  $D=5$ and $D=6$ dimensions map naturally to
Dirac fermions in $D=4$ in a vector-like representation of $U(1)_Y$. 
In the compactified extra dimensions, the boundary conditions are 
taken to be 
\beq
L_D(z_i+2\pi R_i, x)\,=\, e^{i q_{Fi} 2 \pi } L_D(z_i,x)\,,
\label{eq:bcL}
\eeq
where $q_{Fi}$ are arbitrary phase-shifts. Performing the  Fourier 
series expansion in each of the compact $z_i$-coordinates, the bulk field
yields a tower of Kaluza-Klein (KK) states in the four-dimensional description with masses $M_{L, {\mathbf{n}}}$.

For example in the $D=5$ case one has 
\beq
 L_D(z_1,x^\mu)\,=\, \sum_{n=-\infty}^{+\infty} e^{iz_1(n+q_F)/R} \, L_{D}^{(n)}(x^\mu)\,, \quad {\rm with} \quad
M_{Ln}^2\,=\, M_D^2 \,+\, (n + q_F)^2/R^2\,,
 \label{eq:bcL5D}
\eeq
where $M_D$ is the Dirac mass of the leptons (cf. \eqref{eq:Xsc}),
\beq
{\cal L}_{\rm bulk} \,=\, \bar{L}_D\left(i\gamma^\mu D_\mu -M_D\right) L_D
\,+\, y X \bar{L}_D L_D \,.
\label{eq:Xsc2}
\eeq
For higher dimensions we will for simplicity assume degenerate radii, $R_{i=1\ldots d} = R$, and introduce a 
$d$-dimensional vector notation, $\mathbf{n}= (n_1,\ldots,n_d)$ and $\mathbf{q}= (q_{F1},\ldots, q_{Fd})$. We can then write,
\beq
 L_D(z_i,x^\mu)\,=\, \sum_{\mathbf n} \,
 e^{i{\mathbf z}\cdot 
 ( {   {\mathbf n} + {\mathbf q} })/R } \, L_{D}^{({\mathbf n})}(x^\mu)\,,
 \label{eq:bcL6D}
\eeq
with KK masses given by
\beq
M_{L{\mathbf n}}^2\,=\, M_D^2 \,+\, ({\mathbf n} + {\mathbf q})^2 M_c^2 
\,,
\eeq
where $M_c\,\equiv\, 1/R$ is the compactification mass scale.

The $X$ to di-photon interaction  \eqref{eq:EFT} is  generated by summing over the KK modes of the 
vector-like lepton propagating in the loop of the triangle diagram. In the $D=6$ theory,
the Wilson coefficient of the $XFF$ operator is
given by (cf. \eqref{eq:formfactor}),
\beq
\frac{e^2}{2\Lambda_{\rm sc}}\,=\,\, \frac{e^2 Q^2 y}{24 \pi^2} \,\, \sum_{\mathbf n} 
\frac{M_D}{M_{L{\mathbf n}}^2}
\, e^{-(M_{L{\mathbf n}}^2/M_{\rm st}^2)}\,\, 
 {\cal F}\left(\frac{M_X^2}{4M_{L{\mathbf n}}^2 }\right)
\label{eq:KKformfactor}
\,,
\eeq
where the function ${\cal F}$ is the formfactor appearing in \eqref{eq:formfactor2}.

We should remark that this contribution is related to the contribution of the KK modes to the beta function of the gauge coupling 
itself. Therefore one should pay attention to the perturbativity of the latter at some point. Indeed we shall see in the following section 
that the two can be directly related by performing a single beta function calculation. For the moment however we focus on the direct evaluation of the formfactor.

Note that the KK sum has been regulated in \eqref{eq:KKformfactor} with a factor $e^{-(M_{L{\mathbf n}}^2/M_{\rm st}^2)}$; we argue shortly 
and in Appendix A that this regulator is the one that naturally emerges in most string theory calculations. Moreover for the results in 
$D=5$ and $D=6$ the precise form of the regulator will not change the conclusions. 

Note also that the contributions of each individual KK mode to 
the right hand side of Eq. \eqref{eq:KKformfactor} go as ${M_D}/{M_{L{\mathbf n}}^2}$ and not as $1/M_{L{\mathbf n}}$ 
as one might have supposed from the characteristic $1/M_L$ behaviour in \eqref{eq:formfactor}.
In fact is easy to see that the the triangle diagram vanishes in the $M_D \to 0$ limit for any KK occupation number, 
as a consequence of the fact that only the traces of even numbers of Dirac matrices survive so 
at least one power of $M_D$ is required in the numerator of
\beq
\int d^4 p
\frac{{\rm Tr}((\gamma\cdot p_1 + M_D)\gamma^\mu(\gamma\cdot p_2 + M_D)\gamma^\mu (\gamma\cdot p_3 + M_D)}
{(p_1^2+ M_{L{\mathbf n}}^2)(p_2^2+M_{L{\mathbf n}}^2)(p_3^2+ M_{L{\mathbf n}}^2)} \, \sim \,\, k_1^{\mu} k_2^{\nu} \, \frac{M_D}{M_{L{\mathbf n}}^2}\,,
\eeq
where $\gamma\cdot p$ are $D$-dimensional scalar products, and $k_i$ are the external momenta of the photons.

\medskip

For a model in $D$ dimensions we have to evaluate the $d$-fold sum,
\beq
S\, =\,   \sum_{\mathbf{n}} \, \frac{M_D}{M_{L,\mathbf{n}}^{2}}\,\exp(-M_{L,\mathbf{n}}^{2}/M_{\rm st}^{2}) \,,
\quad {\rm with} \quad
M_{L,\mathbf{n}}^{2}\,=\, M_D^2 \,+\, (\mathbf{n}+\mathbf{q})^2 /R^2 \
\label{eq:Sumgen}
\,.
\eeq
It is informative to compare with string calculations, and in particular to understand the motivation of the regularisation in Eq.\eqref{eq:KKformfactor}, 
to use Schwinger parameterisation to write this expression as 
\begin{equation}
S\, =\, \sum_{\mathbf{n}}\int_{0}^{\infty}dt\,\, M_{D}\exp\left[-t\left(M_{D}^{2}+\frac{(\mathbf{n}+\mathbf{q})^{2}}{R^{2}}\right)\right]\,.
\label{eq:Sumgen2}
\end{equation}
Note that for $\mathbf{n}+\mathbf{q}=0$
this integral is simply the usual zero-mode factor $S_{0}=1/M_{D}$.
By inspection, putting a regularization $\exp(-M_{L,\mathbf{n}}^{2}/M_{\rm st}^{2})$
in the KK sum, is precisely equivalent to placing a UV cut-off on this
integral of $t>t_{\rm st}=1/M_{\rm st}^{2}$ . Conversely, it is for this
reason that the automatic regularization that occurs in string theory
due to for example modular invariance excising the UV divergent regions, typically
results in regulating factors $\exp(-M_{L,\mathbf{n}}^{2}/M_{\rm st}^{2})$
appearing in otherwise divergent Kaluza-Klein sums. 
Indeed such regulating factors appear already in tree-level
scattering, because of the ``softening'' effect of string theory
(the fact that D-branes have finite thickness for example), which puts
a  limit on the accessible Compton wavelengths \cite{Antoniadis:2000jv,Ghilencea:2001bv,Abel:2003vv}. We discuss this 
aspect more in Appendix A. 

\subsection{5 Dimensional result:}

\def\bfell{{\boldsymbol \ell}}

In $D=5$ the summation in \eqref{eq:Sumgen} 
is finite resulting in 
\beq
S \,=\,  \frac{\pi R\, \sinh(2\pi M_{D} R)}{\cosh(2\pi M_{D} R)-\cos(2\pi q_{F})}
 \, \approx \, \begin{cases}
\frac{2\pi^{2}M_{D} R^{2}}{1-\cos(2\pi q_{F})}\,\sim \, 2\pi^{2}\frac{M_{D}}{M_{\rm c}^{2}} & :\,\,2\pi M_{D}R\ll1,\,q_{F}\neq0\\
\quad \frac{1}{M_{D}} & :\,\,2\pi M_{D} R\ll1,\,q_{F}=0\,.
\end{cases}
\label{eq:5DS}
\eeq
Regardless of the relative sizes of $M_D$ and $M_c$, there is no significant enhancement over simply taking $1/M_D$ as the indicative value of $S$. 

\subsection{4+d Dimensional result:}

In $D=6$ (and higher) dimensions we expect to find logarithmic (and higher $\sim M_{\rm st}^{D-6}=M_{\rm st}^{d-2}$)
dependence on the UV theory, and we therefore reinstate the $t>t_{\rm st}$ cut-off in \eqref{eq:Sumgen2}. 
Performing Poisson resummation we have
\begin{equation}
S\, =\, M_{D}\int_{t_{\rm st}}^{\infty}dt\,\, R^{d}\left(\frac{\pi}{t}\right)^{\frac{d}{2}}\sum_{\mathbf{\bfell}}
\cos(2\pi\mathbf{\bfell}\cdot\mathbf{q})e^{-M_{D}^{2}t-\mathbf{\bfell}^{2}\pi^{2}R^{2}/t}
\end{equation}
where $\bfell$ denotes the dual integer lattice sum with the same
dimensionality as $\mathbf{n}$. The contributions for non-zero $\bfell$
are negligible, and for the $\bfell=0$ contributions we find 
\begin{equation}
S=\frac{1}{M_{D}}\left(M_{D}R\right)^{d}\pi^{\frac{d}{2}}\Gamma\left[1-\frac{d}{2};\frac{M_{D}^{2}}{M_{\rm st}^{2}}\right].
\end{equation}
Expanding for small $M_{D}\ll  M_{\rm st}$, we find 
\begin{equation}
S=\begin{cases}
\pi\frac{1}{M_{D}}(M_{D}/M_{\rm c})^{2}\left[\log\frac{M_{\rm st}^{2}}{M_{D}^{2}}-\gamma_{E}\right] & :\,\,d=2\\
\frac{2\pi^{\frac{d}{2}}}{d-2}\frac{1}{M_{D}}(M_{D}/M_{\rm c})^{2}\left(M_{\rm st}/M_{\rm c}\right)^{d-2} & :\,\,d>2
\end{cases} \, \qquad {\rm where}\quad M_{\rm c} \,=\, 1/R\,.
\end{equation}
Note that for these approximations to be valid, including the original Poisson resummation, we need only assume that $M_c,\, M_D\ll M_{\rm st}$, but do not need to make any assumption about the relative values of $M_D$ and $M_c$. Moreover it is important to appreciate that the KK tower is preventing large $M_D$ killing the sum since, it is the UV end of the 
tower that dominates the contributions, and this depends solely on the separations $M_c$ regardless of the  Dirac mass. 

We thus find
\beq
\frac{e^2}{2\Lambda_{\rm sc}}\,=\,\, \frac{e^2 Q^2 y}{24 \pi^2} \,\, \frac{1}{M_D} \,\, \frac{M_D^2}{M_{\rm c}^{2}} \times
\begin{cases}
\pi\left[\log\frac{M_{\rm st}^{2}}{M_{D}^{2}}-\gamma_{E}\right] & :\,\,d=2\\
\frac{2\pi^{\frac{d}{2}}}{d-2}\left(M_{\rm st}/M_{\rm c}\right)^{d-2} & :\,\,d>2\, .
\end{cases}
\label{eq:KK4}
\eeq 
 We will now trade the the 4D Dirac mass $M_D$ for the mass of the lightest KK mode $L_D^{(\mathbf{0})}$ via,
 \beq
 M_{L,\mathbf{0}} \,=\, \sqrt{M_D^2\,+\, (\mathbf{q})^2 M_{\rm c}^2}
 \,=\, M_D\, \sqrt{1\,+\, (\mathbf{q})^2 M_{\rm c}^2/M_D^2}\,.
 \eeq
  We can now determine the lepton mass
 necessary to reproduce the di-photon signal in terms of $\Lambda\simeq 236$ GeV in our earlier estimate \eqref{eq:estL};
 \beq
M_{L,\mathbf{0}} \,=\, \frac{\Lambda}{12\pi^2}\, Q^2 y\,
\left(1\,+\, (\mathbf{q})^2 \frac{M_{\rm c}^2}{M_D^2}  \right)^{1/2}
\frac{M_D^2}{M_{\rm c}^2} \times
\begin{cases}
\pi\left[\log\frac{M_{\rm st}^{2}}{M_{D}^{2}}-\gamma_{E}\right] & :\,\,d=2\\
\frac{2\pi^{\frac{d}{2}}}{d-2}\left(M_{\rm st}/M_{\rm c}\right)^{d-2} & :\,\,d>2\, .
\end{cases}
 \label{eq:fin1}
 \eeq
 
 \begin{figure}[t]
\begin{center}
\begin{tabular}{cc}
\hspace{-.4cm}
\includegraphics[width=0.43\textwidth]{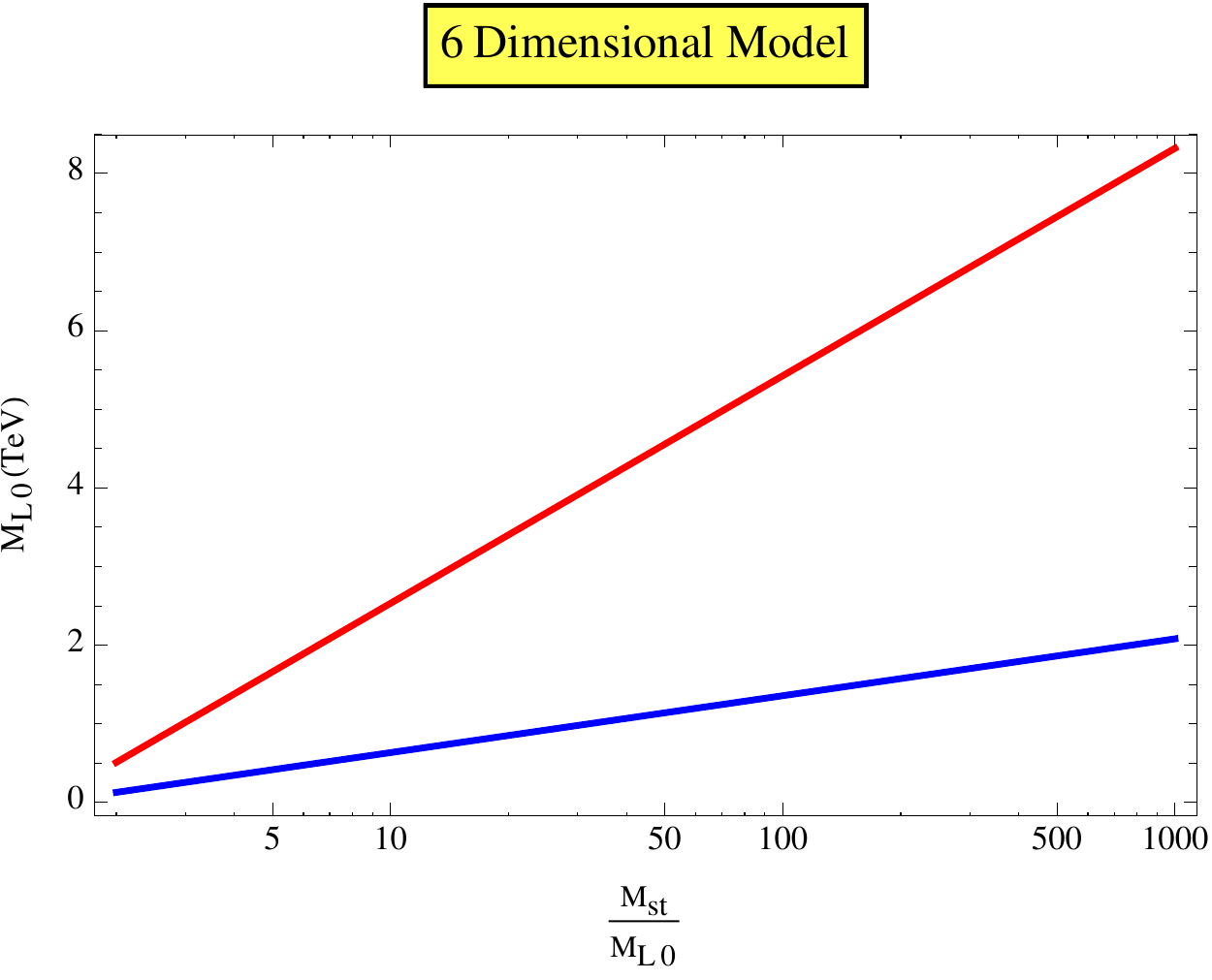}
&
\hspace{.4cm}
\includegraphics[width=0.56\textwidth]{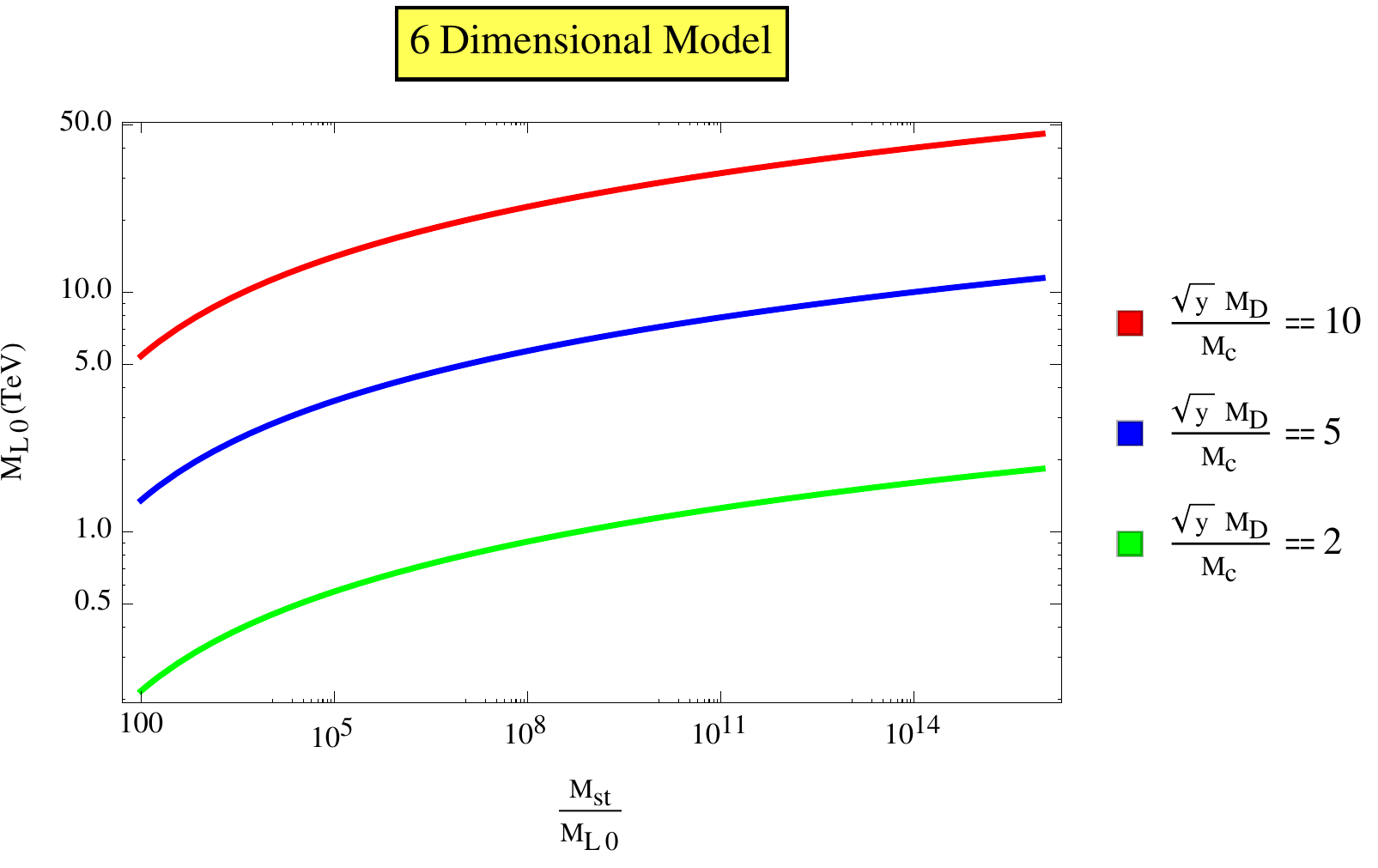}
\\
\end{tabular}
\end{center}
\vskip-.4cm
\caption{
Mass  $M_{L,\mathbf{0}}$ of the lowest KK mode (in TeV), with 2 extra dimensions \eqref{eq:fin1_2}
that is able to generate the di-photon signal via photon fusion,
shown as the function of $M_{\rm st}/M_{L,\mathbf{0}}$. 
The ratios of Dirac lepton mass to compactification scale are taken to be $\sqrt{y} M_D/M_{\rm c}=10,\, 5,\, 2$. The plot on the left zooms into the low mass region.}
\label{fig:one}
\end{figure}
 
 First we consider the case of $d=2$. Here the dependence on the UV cutoff, and hence the 
 effective number of KK modes contributing, is only logarithmic. The overall contribution in this case is boosted 
 when $M_D$ becomes greater than the 
 compactification mass scale (typically we will consider  $M_D/M_{\rm c} \sim 10$); this also implies $M_{L,\mathbf{0}} \,\simeq M_D$.
 From the first equation in \eqref{eq:fin1} we obtain
\beq
d=2: \,\,
M_{L,\mathbf{0}} = 2 \pi \frac{\Lambda}{24\pi^2}\, Q^2 y\,
\frac{M_D^2}{M_{\rm c}^2} 
\left(\log\frac{M_{\rm st}^{2}}{M_{D}^{2}}-\gamma_{E}\right)
  \simeq\,
 \, 2 \pi \, Q^2 y\,
\frac{M_{L,\mathbf{0}}^2}{M_{\rm c}^2} 
\left(\log\frac{M_{\rm st}^{2}}{M_{L,\mathbf{0}}^{2}}-\gamma_{E}\right) \times [1 \,{\rm GeV}]\,,
   \label{eq:fin1_2}
 \eeq
 where 
on the right hand side we have set $\Lambda/(24\pi^2)=  236 /(24\pi^2)\simeq 1$ GeV using Eq.\eqref{eq:estL}, and we have traded 
$M_D$ for $M_{L,\mathbf{0}}$ in the mass ratios.

 In Figure~\ref{fig:one} we plot the resulting values of the lowest vector-like lepton mass $M_{L ,\mathbf{0}}$  as a function of the
 string scale measured in units of the lepton mass. 
Postponing the issue of where the coupling constants may become strong/non-perturbative to the next section,
from this Figure and the data in Table~\ref{tab:table1} we can infer that for a string scale in $D=6$ 
varying between, for example, 15~TeV and $10^{10}$~GeV, the lightest
 lepton masses lie between
 $500$ GeV and $25$ TeV (assuming $M_{L\mathbf{0}}=10 M_{\rm c} $). 
  
 In fact, for even lower values of the compactification mass, i.e. below the threshold for a 2-particle decay, $ M_X/2$,
  the di-photon resonance $X$  could decay into new states propagating in the extra dimensions (if these states are present 
  in addition to the heavy vector-like leptons $L$). This effect would contribute to the remaining 97\% of the
  total 45 GeV width in addition to the 3\% accounted for by di-photon decays \eqref{eq:bound}.
 
 \begin{table}[h!]
  \centering
    \begin{tabular}{c|cccccccc}
    \toprule
     $M_{\rm st}$ (TeV) & ~ 15 ~& ~ 30 ~& ~ 100 ~ & ~ 400 ~ & ~ $4\cdot 10^4$ ~& ~ $3\cdot 10^6$ ~ & ~ $2.5\cdot10^8$ ~ &  ~ $10^{10}$ \\
    \midrule
   $M_{L,\mathbf{0}}$ (TeV) & ~ 0.5 ~& ~ 1 ~& ~ 3.76 ~ & ~ 5.11 ~ & ~ 10 ~& ~ $15$ ~ & ~ $20$ ~ &  ~ $24.5$  \\
    \bottomrule
  \end{tabular}
  \caption{
  Values of the lepton mass $M_{L\mathbf{0}}$ vs the string scale in the model with $d=2$ extra dimensions compactified 
  on a torus with degenerate radii $R=1/M_{\rm c}$ following Eq.~\eqref{eq:fin1_2}.
  We have chosen $M_{\rm c} = 0.1 M_{L\mathbf{0}}$ and set the yukawa coupling to $y=1$. The two left-most values maintain perturbativity even under 
  the most minimal assumptions, see section {\bf \ref{sec:P}}.}
  \label{tab:table1}
\end{table}

\bigskip
 
 \begin{figure}[t]
\begin{center}
\includegraphics[width=0.6
\textwidth]{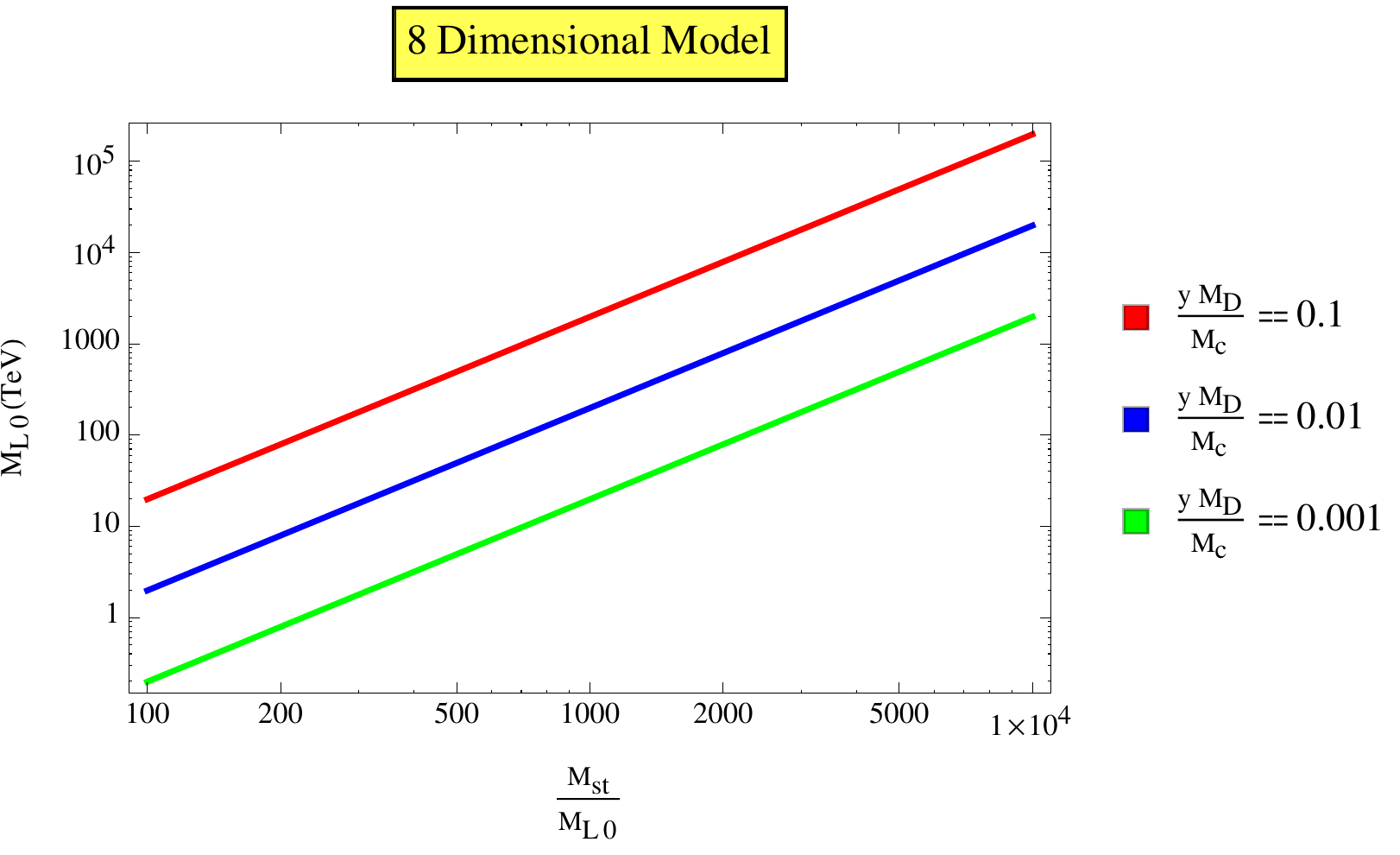}
\end{center}
\vskip-.4cm
\caption{Values of 
the lepton mass $M_{L,\mathbf{0}}$ of the lowest KK mode (in TeV) in the model with 4 extra dimensions \eqref{eq:fin2_4}
which generates the di-boson signal via photon fusion,
shown as the function of the $M_{\rm st}/M_{L,\mathbf{0}}$. We indicate different values of the
yukawa constant and the Dirac mass to the compactification mass scale
ratio, $y M_D/M_{\rm c}=0.1,\, 0.01,\, 0.001$.}
\label{fig:two}
\end{figure}

 Finally, as an example of $d>2$ we discuss the $d=4$ (i.e. $D=8$) case, where we can 
consider the opposite limit $M_D \ll M_{\rm c}$ to that required for $D=6$.
 In this case, setting all $\mathbf{q} = 1/2 $ for concreteness, we have (recalling that $d=4$) 
 \beq
M_{L,\mathbf{0}}\, \simeq\, \sqrt{(\mathbf{q})^2 }M_{\rm c}
 \,=\, M_{\rm c}\,.
 \eeq
According to Eq.~\eqref{eq:fin1} the values of $M_{L,\mathbf{0}}$ required to explain di-photon resonance production are then
 \beq
d=4: \quad
M_{L,\mathbf{0}} 
 \, \simeq\,\,
 2\pi^2 Q^2\,  \left(y\,
\frac{M_D}{M_{L,\mathbf{0}}}\right)\times 
\frac{M_{\rm st}^{2}}{M_{L,\mathbf{0}}^{2}}\times [1 \,{\rm GeV}]\,.
   \label{eq:fin2_4}
 \eeq
This is plotted in Fig.~\ref{fig:two} for $y M_D/M_{\rm c}=0.1,\, 0.01,\, 0.001$. 
Equations \eqref{eq:fin1_2} and \eqref{eq:fin2_4} illustrated in the accompanying figures are the main results of this section.
 
\medskip
  
\section{Constraints from perturbativity }
\label{sec:P} 

The KK states also contribute to the running of the gauge coupling,
therefore one should also address the scale at which these effects
may make the coupling strong. As one might expect this will put a
bound on the string scale with respect to the KK scale (essentially
a cap on the total number of KK states), and in this section we briefly
estimate it. We will continue to use the Schwinger type of analysis
which is most closely related to the string theory results, where
broadly speaking one expects power-law running type contributions
between the fundamental scale and the KK scale. Before starting we
should note that one has to be aware of the various subtleties in mapping
extra-dimensional field-theory to string theory (see for example Ref.\cite{Ghilencea:2001bv}).
 This is really a re-rendering
of the so-called decompactification problem, and conceivably there
are theories that can evade it, perhaps with higher dimensional gauge unification with fixed points as in Ref.\cite{Dienes:2002bg}, 
or in a stringy setting as in reviewed recently in Ref.\cite{Partouche:2016xqu}. All of these possibilities may come in to play at some energy scale. 
Therefore the discussion here is conservative: more precisely one could replace ``the string scale'' with the scale of some UV cut-off above which 
new physics changes the running of the gauge coupling.  

To estimate the running associated with the KK enhancement of the
di-photon rate, we repeat the computation of the $XF^{\mu\nu}F_{\mu\nu}$
coupling by extracting from the renormalization of the gauge coupling
itself, with $X$ as a background field. The quantity we need is the
vacuum polarisation written as 
\begin{equation}
\Pi(\mu^{2}<M_{L,\mathbf{0}}^{2})=\frac{e^{2}}{24\pi^{2}}\sum_{\mathbf{n}}\int_{0}^{\infty}\frac{dt}{t}e^{-M_{\mathbf{n}}^{2}t},
\end{equation}
which Poisson resummed gives 
\begin{equation}
\Pi(\mu^{2})=\frac{e^{2}}{24\pi^{2}}M_{\rm c}^{-d}\pi^{d/2}\int_{\ell_{\rm st}^{2}}^{\infty}\frac{dt}{t^{1+d/2}}e^{-\mu^{2}t/\pi}\sum_{\bfell}\cos(2\pi\bfell\cdot q)e^{-M_{D}^{2}t-\bfell^{2}\pi^{2}/\left(M_{\rm c}^{2}t\right)}.
\end{equation}
Ultimately we may simply replace $M_{D}\rightarrow M_{D}+yX$ in order
to also get the previously obtained coupling between the photons and the diboison resonance $X$. The integral is UV divergent so again
we place a UV cut-off $\ell_{\rm st}^{2}=1/M_{\rm st}^{2}$, and find
\begin{equation}
\Pi=\frac{e_{\rm st}^{2}}{24\pi^{2}}\left(\frac{M_{D}}{M_{\rm c}}\right)^{d}\pi^{\frac{d}{2}}\Gamma\left[-\frac{d}{2};\frac{M_{D}^{2}}{M_{st}^{2}}\right]\,,
\end{equation}
where $e_{\rm st}$ is the tree-level value of the gauge coupling. Neglecting
the additional contribution of logarithmic running from the massless
spectrum, we identify 
\begin{eqnarray}
\frac{16\pi^{2}}{e^{2}(\mu^{2})} & = & \frac{16\pi^{2}}{e_{\rm st}^{2}}+\frac{16\pi^{2}}{e_{\rm st}^{2}}\Pi\,,\nonumber \\
 & = & \frac{16\pi^{2}}{e_{\rm st}^{2}}+\frac{2}{3}\left(\frac{M_{D}}{M_{\rm c}}\right)^{d}\pi^{\frac{d}{2}}\Gamma\left[-\frac{d}{2};\frac{M_{D}^{2}}{M_{st}^{2}}\right]\,.
\end{eqnarray}
We may then expand for $M_{D}\ll M_{\rm st}$
\begin{equation}
\frac{16\pi^{2}}{e^{2}}=\frac{16\pi^{2}}{e_{\rm st}^{2}}+\frac{2}{3}\pi^{\frac{d}{2}}\begin{cases}
\left(\frac{M_{\rm st}}{M_{\rm c}}\right)^{d}+\left(\frac{M_{D}}{M_{\rm c}}\right)^{2}\left(\gamma_{E}-1-\log M_{st}^{2}/M_{D}^{2}\right) & ;\,\,d=2\\
\frac{2}{d}\left(\frac{M_{\rm st}}{M_{\rm c}}\right)^{d}-\frac{2}{d-2}\left(\frac{M_{\rm st}}{M_{\rm c}}\right)^{d-2}\left(\frac{M_{D}}{M_{\rm c}}\right)^{2} & ;\,\,d>2.
\end{cases}
\end{equation}
and then replacing $M_{D}\rightarrow M_{D}+yX$ and expanding in $X$
gives
\begin{equation}
\frac{16\pi^{2}}{e^{2}(\mu^{2}<M_{L,\mathbf{0}}^{2})}=\frac{16\pi^{2}}{e_{\rm st}^{2}}+\frac{2}{3}\pi^{\frac{d}{2}}\begin{cases}
\left(\frac{M_{\rm st}}{M_{\rm c}}\right)^{d}+2\frac{yX}{M_{D}}\left(\frac{M_{D}}{M_{\rm c}}\right)^{2}\left(\gamma_{E}-\log M_{st}^{2}/M_{D}^{2}\right)+\ldots & ;\,\,d=2\\
\frac{2}{d}\left(\frac{M_{\rm st}}{M_{\rm c}}\right)^{d}-\frac{4}{d-2}\frac{yX}{M_{D}}\left(\frac{M_{\rm st}}{M_{\rm c}}\right)^{d-2}\left(\frac{M_{D}}{M_{\rm c}}\right)^{2}+\ldots & ;\,\,d>2.
\end{cases}
\end{equation}

 \begin{figure}[t]
\begin{center}
\begin{tabular}{cc}
\hspace{-.4cm}
\includegraphics[width=0.5\textwidth]{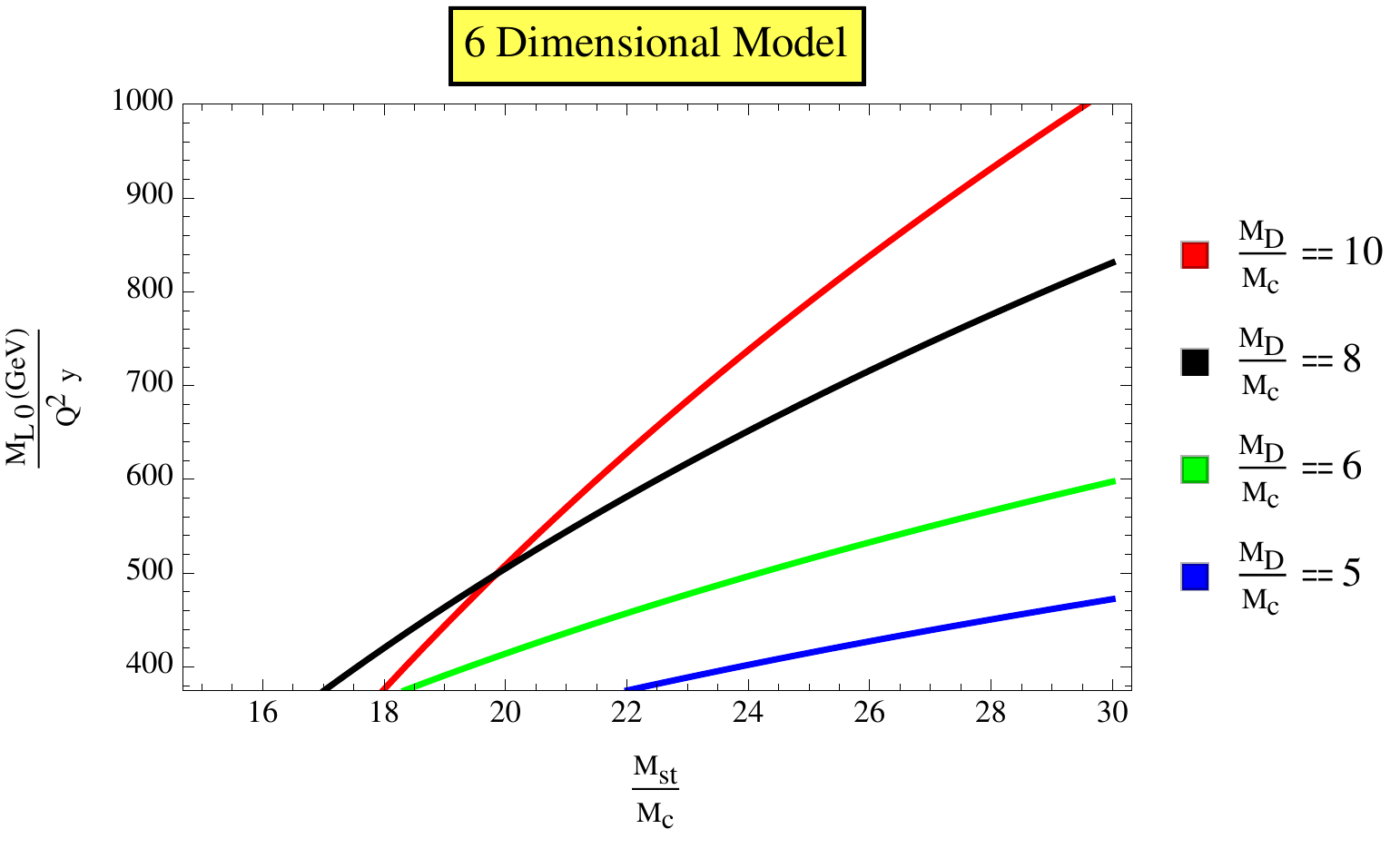}
&
\includegraphics[width=0.5\textwidth]{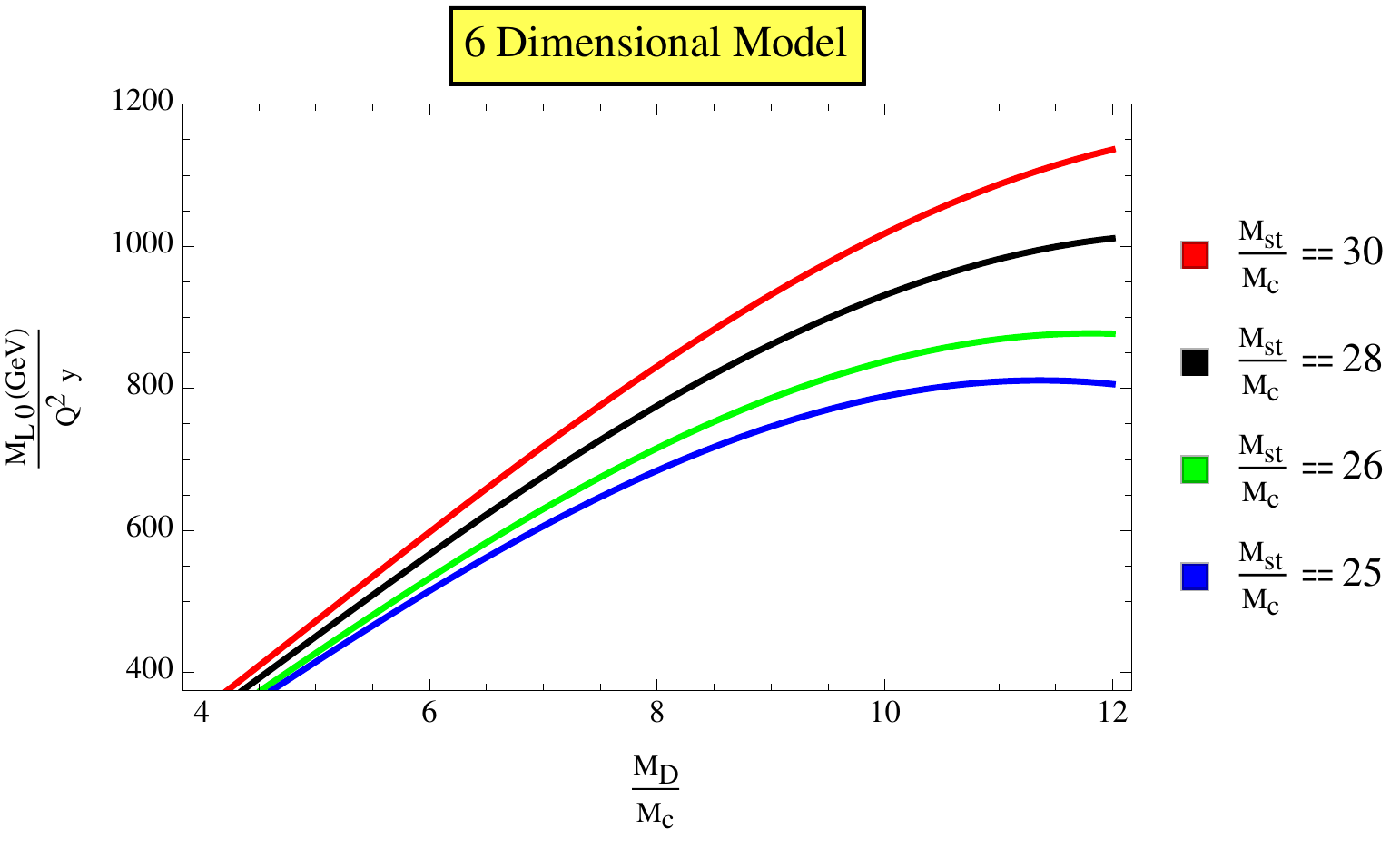}
\\
\end{tabular}
\end{center}
\vskip-.4cm
\caption{
Values of 
the lepton mass $M_{L,\mathbf{0}}$ of the lowest KK mode (in units of $Q^2 y$ GeV)
in the model with 2 extra dimensions satisfying perturbativity constraints \eqref{eq:pert}.
On the left panel we vary $M_{\rm st}/M_{\rm c}$ ratio within the range allowed by perturbativity $\lesssim 30$.
On the right panel we very the second ratio $M_{D}/M_{\rm c}$.
}
\label{fig:oneP}
\end{figure}

It can be checked that this yields precisely the previous result \eqref{eq:KK4} for
the $XF^{\mu\nu}F_{\mu\nu}$ coupling, as a term in addition to a
power-law contribution to $1/e^{2}$. Clearly then the KK enhancement
is associated with an enhanced gauge beta function. For perturbativity
we should also then ensure 
\begin{equation}
\frac{4}{3d}\pi^{\frac{d}{2}}\left(\frac{M_{\rm st}}{M_{\rm c}}\right)^{d}\lesssim \frac{16\pi^{2}}{e^{2}(\mu^2_{\rm IR})}.
\end{equation}
In other words the best one can do is to take $e_{\rm st}$ to represent
a Landau pole at the scale $M_{\rm st}$; numerically this translates
as $\frac{4}{3d}\pi^{\frac{d}{2}}(M_{\rm st}/M_{\rm c})^{d}\lesssim16\pi^{2}/e^{2}=4\pi\times137$,
which implies 
\begin{equation}
\frac{M_{\rm st}}{M_{\rm c}}\lesssim\begin{cases}
30 & \,\,;\,\,d=2\\
5 & \,\,;\,\,d=4\,.
\label{eq:pert}
\end{cases}
\end{equation}
In general the naive assumption that KK modes contribute to the gauge
coupling requires a string scale one or perhaps two orders of magnitude
higher than the compactification scale. 

We can now re-write the equation \eqref{eq:fin1_2} for the model with $d=2$ extra dimensions in the form,
\beq
d=2: \quad
\frac{1}{Q^2\, y}\, M_{L,\mathbf{0}} =  \, 4 \pi \,
x^2
\left(\log z -\log x -\gamma_{E}/2\right) \times [1 \,{\rm GeV}]\,,
   \label{eq:FIN2}
 \eeq
 where we have defined the two mass ratios,
 \beq
 x= M_D/M_{\rm c} \quad {\rm and} \quad z= M_{\rm st}/M_{\rm c}
 \eeq
In Figure~\ref{fig:oneP} we plot the values of $M_{L,\mathbf{0}}$ resulting from \eqref{eq:FIN2} (as required to fit the di-photon production rate) 
now taking into account  the perturbativity constraint \eqref{eq:pert}. We see that even in this class of the relatively low-cutoff perturbative
models, one can obtain $M_{L,\mathbf{0}}$ in the range from above the $375$ GeV to up to 1200 GeV and for string scales $M_{\rm st}$ up to $\sim 3$ TeV.

Now in the model with $d=4$ extra dimensions we can the second equation in \eqref{eq:fin1} (see also \eqref{eq:fin2_4}) in the form,
\begin{eqnarray}
M_{L,\mathbf{0}} &=&  M_{\rm c}\, \sqrt{x^2+1} \,, \label{eq:fin4ML0}
\\
d=4: \quad \frac{1}{Q^2\, y} M_{\rm c} &=&  2\pi^2 \, z^2 \times [1 \,{\rm GeV}] \, <\,  450 - 500 \, {\rm GeV}\,,
   \label{eq:fin4Mc}
 \end{eqnarray}
where the bound on the right hand side of \eqref{eq:fin4Mc} comes from the perturbativity bound $z<5$ in the second equation in \eqref{eq:pert}.
This amounts to the UV-cutoff scale bound $M_{\rm st} < 2 - 2.5$ TeV. 
Finally, for the the zero mode of the KK lepton in the eight-dimensional theory one can choose any value in the range between
$M_{L,\mathbf{0}} \simeq M_{\rm c}$ and $M_{\rm c}\, \sqrt{x^2+1}$ where $x< 5$. For example, for $x\simeq 3$, the 
lowest KK lepton modes are in the range of 1.5 TeV (times $Q^2 \, y$).

 \medskip
 
 \section{Finite naturalness for the di-photon mass $M_{X}$}
\label{sec:MX} 

Our approach so far has been based on the idea that the high multiplicity of KK states of the vector-like leptons
can significantly enhance their 1-loop contributions to the coupling of the spin-zero $X$ field to two photons.
We must be careful, however, to ensure that the same tower of KK leptons does not at the same time introduce 
unwanted large effects elsewhere. 

The most obvious and potentially most dangerous such effect would be an
exploding contribution to the 750 GeV mass of the $X$-resonance itself. 
Indeed, the vector-lepton contributions to the self-energy of $X$ have two (rather than three, which was the case before) 
propagators in the loop and produce relevant (i.e. mass dimension-2)  operators. UV-cutoff-dependent contributions 
$\Delta M_X^2 \propto \, y^2/(16 \pi)^2\, M_{\rm st}^2$ would be very unwelcome.
Such contributions can however be removed by a `lepton-partners' mechanism, as we will now outline.

A simple implementation of such a mechanism begins with a supersymmetric Wess-Zumino model in $D=6$, with
superpotential,
\beq
{\cal W}\,=\, y X \tilde{L}L\,+\, M_{D}\tilde{L}L\,,
\label{eq:W}
\eeq
where $L$ and $\tilde{L}$ are two chiral superfields, which can be thought of as forming an ${\cal N}=2$ hypermultiplet.
As before, the fermionic components of  $L$ and $\tilde{L}$ form a single Dirac multiplet, but also of course provide scalar
lepton partners. Imposing boundary conditions as in Eq.~\eqref{eq:bcL} with $\mathbf{q}_B\neq \mathbf{q}_F$  
(for bosons and fermions respectively) breaks supersymmetry spontaneously by the Scherk-Schwarz mechanism \cite{Scherk:1979zr},
for early string theory realisation see  \cite{Antoniadis:1990ew}.
Typically $\mathbf{q}_B$ and $ \mathbf{q}_F$ are taken to have anti-periodic or periodic entries. 

At 1-loop level the contribution to the $X$-boson mass from the KK towers of the lepton and lepton partner states is,
\beq
\Delta M_X^2 \,=\, y^2 \sum_{\mathbf{n}}
\int \frac{d^4p}{(2\pi^4)}\left(
\frac{1}{p^2 + (\mathbf{n} + \mathbf{q}_B)^2 M_{\rm c}^2 + M_D^2}\,-\, 
\frac{1}{p^2 + (\mathbf{n} + \mathbf{q}_F)^2  M_{\rm c}^2  + M_D^2}\, .
\right)
\label{eq:susyKK}
\eeq
The final expression is known to be entirely finite thanks to the spontaneous nature of the breaking \cite{Antoniadis:1998sd}.
Indeed one can easily calculate the mass by again Poisson resumming the Schwinger integrals (as in Appendix B), to find
\beq
\Delta M_X^2 \,=\, 
\begin{cases}
\,\,\, 0  & :\,\,
\mathbf{q}_B \,=\, \mathbf{q}_F\\
y^2 \, M_{\rm c}^2
 \frac{\Gamma[1+d/2]}{8\pi^{4+d/2}}\xi(1+d/2)
  & :\,\,
\mathbf{q}_B =0 \quad {\rm and}\quad \mathbf{q}_F=1/2\, .
\end{cases}
 \eeq
 In the first case boundary conditions do not break supersymmetry and there are no
 radiative corrections to the mass of the di-photon resonance at the order $\alpha_{\rm em}^0$, i.e. when we neglect the 
 back reaction from the Standard Model.
 In the second case, where supersymmetry in the bulk theory is broken by $\mathbf{q}_B \,\neq\, \mathbf{q}_F$, the finite
 naturalness for the di-photon resonance requires that
 \beq 
 750\, {\rm GeV} \lesssim \, y \,\left(\frac{\Gamma[1+d/2]}{8\pi^{4+d/2}}\xi(1+d/2) \right)^{1/2}\, M_{\rm c} \,\,\simeq\,
 \begin{cases}
0.05\,y\,  M_{\rm c}  & :\,\, d=2\\
0.08\,y\,  M_{\rm c}  & :\,\, d=4\, ,
\end{cases}
 \eeq
where $\xi(s)$ is a multidimensional sum derived in Appendix B: its first few values are $\xi(3/2)=7\zeta(3)/4=2.1$ , $\xi(2)=6.6$,
$\xi(5/2)=14.0$, $\xi(3)=24.4$. This is only a mild constraint on the size of the compactification mass, or equivalently the lightest KK mode of the lepton.

\medskip 

It is also interesting to estimate the back reaction 
on $M_X$ of the SUSY-breaking scale implicit the Standard Model.
The leading-order contribution is the two-loop effect, similar to sfermion masses in gauge mediation.
In our case $X$ directly interacts only with the lepton multiplets $L,\tilde{L}$ which then interact with the
$U(1)_Y$ photon-photino loop. Thus we have a rough estimate
\beq
 \Delta M_X^2 \,\sim\, \frac{y\, \alpha_{\rm em}}{16\pi^2} \, m^2_{\tilde{\gamma}}\,,
\eeq
where $m_{\tilde{\gamma}}$ is the photino mass in the Standard Model. The finite naturalness bound for the 750 GeV
scalar would then imply an upper bound on the photino mass of
\beq
m_{\tilde{\gamma}} \lesssim  (4\pi /\sqrt{y \alpha_{\rm em}} )\, M_X \,\simeq\,
(1/\sqrt{y}) \,\, 110 \, {\rm TeV}\,,
\eeq
which should not cause any problems with the current limits.

\medskip

Let us also comment on the soft masses of the scalar leptons $L_{\rm sc}$.
These are SUSY-breaking contributions first generated at two loops in $\alpha_{\rm em}$ by back reaction from the Standard Model.
Analogously to ordinary slepton mass-squared soft terms in gauge-mediated supersymmetry breaking, we can parameterise
these contributions as,
\beq
\Delta M_{L_{\rm sc}}^2 \,\sim\, \frac{ \alpha_{\rm em}^2}{4\pi^2} \, m^2_{\rm soft\, SM}\,,
\eeq
where $m_{\rm soft\, SM}$ are the slepton masses in the Standard Model. Even at 100 TeV a hypothetical SM slepton would
result only in a $\sim 58$ GeV contribution compared to the mass of $L_{\rm sc}$, which does not impose any significant
phenomenological constraints.

\medskip

\section{Decays of $X$ to dark matter sectors}
\label{sec:dark}

Having generated the required $\simeq 3$\% branching ratio of $X$ to $\gamma\gamma$ necessary to
explain the di-photon excess observed by ATLAS and CMS in terms of solely  photon-photon fusion and decay,
both mediated by a tower of KK states of vector-like leptons, we now address the remaining 97\%
of the 45 GeV total width of $X$. 

An obvious and interesting possibility is that the di-photon spin-zero state $X$ decays into dark matter particles.
More generally $X$ can decay invisibly into any particles of the dark sector, which includes but is not necessarily limited to 
cosmologically stable dark matter. Over the last two years so-called simplified models describing scalar and pseudo-scalar mediators to dark sectors and their searches at the LHC \cite{Buckley:2014fba,Harris:2014hga}
and future hadron colliders \cite{Harris:2015kda} have attracted a fair amount of interest. The role of the 750 GeV resonance 
as the possible mediator to DM was studied in the recent work, including Refs.~\cite{Mambrini:2015wyu,Backovic:2015fnp,Han:2015cty,Bi:2015uqd,Bauer:2015boy,Barducci:2015gtd,Park:2015ysf,Huang:2015svl,Berlin:2016hqw,D'Eramo:2016mgv}.

The simplified models 
describing elementary interactions of spin-zero mediators with the the dark sector 
particles, which for simplicity we take to be Dirac fermions $\chi$, $\bar{\chi}$,
contain the interactions
\begin{align}
\label{eq:LS} 
\mathcal{L}_{\mathrm{scalar}}&\supset\,  - g_{\rm DM}  X \, \bar{\chi}\chi
  - m_{\rm DM} \bar{\chi}\chi \,,
 \\
 \label{eq:LP} 
\mathcal{L}_{\rm{pseudo-scalar}}&\supset\,  - i g_{\rm DM}  X \, \bar{\chi} \gamma^5\chi
  - m_{\rm DM} \bar{\chi}\chi\,.
\end{align}
The partial decay widths of $X$ into these fermions (for a single flavour) are given by
\beq
\label{eq:GS}
\Gamma_{\chi\overline{\chi}} \,=\, \frac{g^2_{DM}}{8\pi }\,M_{X}\,\left(1-\frac{4m_f^2}{M_{X}^2}\right)^\frac{3}{2}
\quad ,\quad
\Gamma_{\chi\overline{\chi}} \,=\, \frac{g^2_{DM}}{8\pi}\,M_{X}\,\left(1-\frac{4m_f^2}{M_{X}^2}\right)^\frac{1}{2}\,,
\eeq
for the scalar and the pseudo-scalar cases respectively.

It is straightforward to achieve 97\% of the 45 GeV $\Gamma_{\rm tot}$  from these decays alone.
For example for $N_f$ light DM fermions one would need
\beq
N_f^{1/2} g_{DM} \,\simeq\, 1.21\,,
\eeq
which is easy to accommodate in a weakly coupled theory. 
It is also possible, if so desired,  to arrange for DM particles to  propagate in large
extra dimensions assuming that the lowest KK modes are kinematically accessible to the 750 GeV particle.

Three comments are in order. First, following the same logic as in Refs.~\cite{Buckley:2014fba,Harris:2014hga,Harris:2015kda},
we will not insist that the relic density of the dark sector fermions computed
using the Lagrangian in Eq.~\eqref{eq:GS} should match the observed cosmological abundance of DM. Such calculations 
have already been carried out elsewhere, e.g. in  Refs.~\cite{Mambrini:2015wyu,Backovic:2015fnp,Han:2015cty,Bi:2015uqd,Bauer:2015boy,Barducci:2015gtd,Park:2015ysf,Huang:2015svl,Berlin:2016hqw,D'Eramo:2016mgv},
but more importantly, it is also possible that the dark sector particles $\chi$ are only the distant parents of the cosmologically
stable DM and can undergo further decays.

Our second comment is that for searches of these dark sectors at the LHC which are mediated by the $X$ particle produced 
via  photon fusion, the standard strategy of jets plus missing energy searches used in
 \cite{Buckley:2014fba,Harris:2014hga,Harris:2015kda} will not be directly applicable. The reason is that at the partonic level
 the process $\gamma\gamma \to X\to \chi\overline{\chi}$ is entirely electromagnetic and we expect that additional QCD jet 
 activity will be suppressed numerically (as already mentioned in Section {\bf 2}).
 
 Finally, one should ensure that the new heavy leptons do not themselves introduce stable charged matter that violates cosmological
 and astrophysical bounds. This requires either sufficiently low reheat temperatures, small freeze-out densities or perhaps decay through higher dimensional 
 operators. See for example refs.\cite{Ishiwata:2013gma,Altmannshofer:2013zba,Falkowski:2013jya}.

\medskip

\section{Summary and Conclusions}
\label{sec:conc}

We proposed a scenario where the spin-zero 750 GeV state is coupled to a new heavy lepton 
which lives in the bulk of a higher-dimensional theory and interacts only with the photons of the Standard Model.
We found that they allow for a minimal and compelling explanation of the di-photon resonance 
via photo-production and decay. The central role in this effect is played 
by the summation over the Kaluza-Klein modes of these leptons appearing in the loops of the $X\to \gamma\gamma$ 
production and decay subprocesses. 

\medskip

The set-up requires only a minimal extension of the Standard Model in the sense 
that the same mechanism is used for the production and the decay of the $X$ resonance.
With the new lepton being coupled only to the $U(1)_Y$ gauge sector in the Standard Model 
one can explain the absence of other resonances at 8 and 13 TeV in the vicinity of the 750 GeV di-photon invariant mass.
The decay channel into $ZZ$ is suppressed by  $\sin^2 \theta_{\rm w}$, while other potential candidates, such as two jets,
$WW$ and 2 leptons are either absent or suppressed by powers of $\alpha_{\rm em}$.

\medskip

This feature is particularly important for suppressing Standard Model jets which would otherwise accompany the di-photon
production mediated by $X$. Since at the parton level the entire $\gamma\gamma \to X \to \gamma\gamma$ process is
electromagnetic, any additional QCD jet activity would be suppressed by extra powers of $\alpha_{\rm em}$
or a combination of ${\cal C}_{q\bar{q}} \, \alpha_{\rm em}^2$ (for the VBF process) relative to the ${\cal C}_{\gamma\gamma}$ factor
in our leading order photon fusion process.
Thus the relative absence of additional jets which would accompany the di-photon resonance if it was produced in gluon fusion 
relative to the case of pure photo-production considered here,  is a distinguishing feature of our model. 

Similarly, while mono-jet searches disfavour a large 
invisible width $\Gamma_{\rm tot} \sim 45$ GeV, this conclusion \cite{Falkowski:2015swt,Barducci:2015gtd} 
does not apply directly to our model,
as the mono-jet would have to originate e.g. from a quark in one of the initial protons $p \to q \to \gamma +q$ in addition to the photon
which participates in the photon fusion. This is again suppressed by the fine structure constant times the corresponding ratio of pdfs.

\medskip

At the same time, the approach presented here can be easily applied to the case where the di-photon (pseudo)-scalar $X$ is produced 
in the gluon fusion channel. In this case one would simply substitute the vector-like lepton by a single species of a new quark $Q$, $\tilde{Q}$ in the bulk of an
extra-dimensional theory. The di-photon rate in the gluon fusion process would of course be greater relative to the photon fusion, and will be 
even easier to fit in our KK model.

\medskip

The extension of the Standard Model by a new spin-zero singlet state $X$ also sits very well with other items
on the BSM wish list:
$X$ can play the role of the inflaton -- as the singlet degree of freedom which is non-minimally coupled to gravity --
see e.g.~\cite{Salopek:1988qh,Hertzberg:2010dc,Khoze:2013uia}.
It can also help to stabilise the SM Higgs potential
\cite{EliasMiro:2012ay,Lebedev:2012zw,Khoze:2014xha,Salvio:2015jgu}, 
assist the first order phase transition and provide additional
sources of CP violation  for  baryogenesis.
Finally as was already noted in Section~{\bf \ref{sec:dark}}, 
a (pseudo)-scalar $X$ is also an obvious candidate for being a mediator to the dark matter sector 
\cite{Mambrini:2015wyu,Backovic:2015fnp,Han:2015cty,Bi:2015uqd,Bauer:2015boy,Barducci:2015gtd,Park:2015ysf,Huang:2015svl,Berlin:2016hqw,D'Eramo:2016mgv}.

\section*{Acknowledgements} 
We are grateful to Valery Khoze, Alberto Mariotti, Misha Ryskin  and Michael Spannowsky for useful discussions and comments.
This work is supported by the STFC through the IPPP grant, 
and for VVK in part by the  Royal Society Wolfson Research Merit Award.

\bigskip

\startappendix

\section*{Appendices}

\section{The string theory context}

The expressions that one derives in the effective field theories are dominated by two elements: the KK tower and the
magnitude and nature of the UV cut-off. In view of the latter, one should check that these
approximations adequately model what happens when a UV completion is added to the theory. 

We have already argued that the physical manifestation of the
UV finiteness of string theory is typically equivalent to a simple
cut-off on Schwinger integrals, and that this in turn is equivalent
to inserting factors $e^{-M_{L{\mathbf n}}^{2}/M_{\rm st}^{2}}$ in otherwise divergent
KK sums. 

However we would like briefly to confirm that this expectation holds
for the Wilson coefficient we are interested in. A simple equivalent
calculation to the field theory one can be performed in the heterotic
string, where the UV cut-off is simply understood as a consequence
of modular symmetry. In general one can find the Wilson coefficient
by obtaining the threshold corrections to $\alpha_{\rm em}$ from the
two-point function of the photon in the standard way,
but inserting the resonance as continuous Wilson line backgrounds
$\mathbf{A}_{i}$ (that break some non-abelian gauge symmetry) around
compact dimensions $i$. The standard starting point is therefore 
\begin{eqnarray}
\Pi^{\mu\nu} & \approx & \frac{g^{2}}{16\pi^{2}}\delta^{ab}(k_{1}^{\mu}k_{2}^{\nu}-k_{1}.k_{2}\eta^{\mu\nu})\int_{\mathcal{F}}\frac{d^{2}\tau}{\tau_{2}}\frac{1}{4\pi^{2}}\sum_{\alpha,\beta}\mathcal{Z}^{\alpha,\beta}(\mathbf{A})\\
 &  & \times\left(4\pi i\partial_{\tau}\log(\frac{\vartheta_{\alpha\beta}(0|\tau)}{\eta(\tau)}\right)\mbox{Tr}\left[-\frac{1}{4\pi\tau_{2}}+Q^{2}\right].\nonumber 
\end{eqnarray}
where $Q$ refers to the charges, and $\alpha,\beta$ refers to spin
structure on the two cycles of the torus\footnote{Typically, to compute running beta functions one would also insert
an IR regulator $e^{-\mu^{2}\tau_{2}}$ where $\mu$ is the $RG$-scale;
of course this is not required for the current problem.}. 

The integration over the real part of the modular parameter, $\tau_{1}=\Re(\tau)$,
projects onto physical states, while the integral over the imaginary
part, $\tau_{2}=\Im(\tau),$ is precisely equivalent to the Schwinger
integral in field theory, with the fundamental domain $\mathcal{F}$
providing a natural cut-off. 

Expanding in the canonically normalized $X$ fields corresponding
to the Wilson-lines, $yX\approx\mathbf{A}_{i}/R_{X}$ where $R_{X}$
is the radius of the dimension associated with Wilson line $X$, and
neglecting the exponentially suppressed string modes as well, we find
\begin{equation}
\mathcal{Z}(\tau)=const-\frac{y}{8\pi^{2}}\frac{R^{d}}{\tau_{2}^{d/2+1}}\sum_{\bfell}\cos(2\pi\ell_{i}\mathbf{q}_{i})\,e^{-\frac{\pi}{\tau_{2}}\ell^{i}G_{ij}\ell^{j}}\ell_{X}^{2}\,R_{X}^{2}M_{D}X+\ldots,
\end{equation}
where the formfactor of interest arises from the $X^{2}$ cross-term
with $\langle y X\rangle=M_{D}$. A non-negligible result requires $R_{X}$
to be small (in other words we didn't need to resum that dimension),
leading to a formfactor of order $M_{D}R^{d}M_{\rm st}^{d-2}$ in agreement
with the field theory. Note that the compactification was somewhat
asymmetric, but the crucial factor was the appearance of the bulk
volume in string units $(RM_{\rm st})^{d}$, which had to be orthogonal
to the Wilson line. 

A similar situation holds in any situation where there is a large
bulk volume dependence in the hypercharge gauge thresholds. For example in type
I or type II models with D-branes, the dependence of threshold corrections
on Wilson lines and/or D-brane displacements has been examined in
the literature in various contexts, for example \cite{Lust:2003ky,Abel:2003ue,Berg:2004ek,Blumenhagen:2006ux,Abel:2008ai}.
The typical dependence goes as 
\begin{equation}
\Pi^{\mu\nu}\approx\frac{g^{2}}{16\pi^{2}}\frac{V_{NN}}{V_{DD}}\delta^{ab}(k_{1}^{\mu}k_{2}^{\nu}-k_{1}.k_{2}\eta^{\mu\nu})\int_{_{0}}^{\infty}dt\,t^{-2}\left(\mbox{const}+\sum_{\bfell}\cos(\bfell\cdot\mathbf{A})\right)e^{-\pi\bfell\cdot G\cdot\bfell/4t}
\end{equation}
where $\mathbf{A}\sim X/R_{X}$ is now the brane displacement, $V_{NN}$
is the volume of Neumann directions, and $V_{DD}$ the volume of Dirichlet
ones. As for the previous example, we can expand for small displacements
and recover a Wilson coefficient proportional to the volume. We should
note in this context that there has been some interest in the 750 GeV di-photon as
a pseudo-scalar axion that participates in generalized Green-Schwarz
anomaly cancellation (i.e. it appears as a closed string R-R state in the Wess-Zumino D-brane action)
\cite{Anchordoqui:2015jxc,Cvetic:2015vit}. That possibility can 
be understood also entirely within an effective field theory in which heavy chiral
modes are integrated out due to a large Yukawa coupling \cite{Anastasopoulos:2006cz}.
However in the present case where we consider the state to be a scalar
coupling to KK modes, it would correspond to a {\it dilatonic} closed
string mode, more along the lines of the scalar proposed in Ref. \cite{Anchordoqui:2015jxc}.

Finally it is interesting in this context naively to estimate parameters under the
assumption that the vector-like leptons and $U(1)_{Y}$ all occupy
the same relatively large volumed D-brane. Then one has $\alpha_{\rm em}=V_{\rm em}\alpha_{\rm st}/V_{\rm st}$
with $V_{\rm em}$ and $V_{\rm st}$ denoting volumes of the respective branes
in string units \cite{Ibanez:1998rf}, leading to an estimate of 
\beq
M_{D} \,\sim \, \frac{\Lambda y}{16\pi^{2}}\frac{\alpha_{\rm st}V_{\rm st}}{\alpha_{\rm em}}\left(\frac{M_{D}}{M_{\rm c}}\right)^{2}
 \, \sim \, \Lambda y\left(\frac{M_{D}}{M_{\rm c}}\right)^{2},
\eeq
assuming $\alpha_{\rm st}V_{\rm st}\sim1$. 

\section{Scalar masses}
\label{sec:app}

Given the possibility of non-zero Scherk-Schwarz contributions to
supersymmetry breaking, it is interesting to ask what contributions
 to the $X$ scalar mass can arise from the Kaluza-Klein lepton tower in \eqref{eq:susyKK}.
 
Like the 5D coupling computed
earlier in \cite{Antoniadis:1998sd} the result is expected to be finite, but in this case for
arbitrary dimensions due to the spontaneous nature of supersymmetry
breaking in Scherk-Schwarz mechanism. As we shall see, the soft-masses are dominated
by the KK modes and consequently the field theory calculation contains
all the necessary physics provided the compactification radii are
relatively large compared to the fundamental scale. (In other words
any string calculation would yield the same result.) 

To see this explicitly, and to obtain precise results for arbitrary
dimensions, let us briefly derive the soft-mass of the scalar in field
theory using the earlier Schwinger approach. The object of interest
is 
\beq
M_{X}^{2} \,=\, y^2\, \sum_{\mathbf{n}}\int\frac{d^{4}p}{(2\pi)^{4}}\left[\frac{1}{p^{2}-(\mathbf{n}+\mathbf{q}_{B})M_{\rm c}^{2}-M_{D}^{2}}-\frac{1}{p_{F}^{2}-(\mathbf{n}+\mathbf{q}_{F})M_{\rm c}^{2}-M_{D}^{2}}\right].
\eeq
Using Schwinger parametrization each integral is of the form
\beq
\sum_{\mathbf{n}}\frac{1}{16\pi^{2}}\int_{0}^{\infty}\frac{dt}{t^{2}}\,\exp\left(-t\left(\frac{(\mathbf{n}+\mathbf{q}_{B})^{2}}{R^{2}}+m_{D}^{2}\right)\right)-\left[B\rightarrow F\right],
\eeq
and Poisson resumming as before we find 
\begin{equation}
M_{X}^{2} \,=\, y^2\, \frac{1}{16\pi^{2}}\int_{0}^{\infty}dt\,R^{d}\pi^{d/2}t^{-(2+d/2)}\sum_{\mathbf{\bfell}}\left[\cos(2\pi\mathbf{\bfell}\cdot\mathbf{q}_{B})-\cos(2\pi\mathbf{\bfell}\cdot\mathbf{q}_{F})\right]e^{-m_{D}^{2}t-\mathbf{\bfell}^{2}\pi^{2}R^{2}/t}.
\end{equation}
The terms with zero $\bfell$ vanish (thanks to the supersymmetry
being spontaneously broken) and the integral is then rendered finite
(hence no UV regularisation is required). Let us assume that $\mathbf{q}_{B}=0$
for every compact radius. Then the result for general $d$ is 
\begin{equation}
\Delta M_{X}^{2}=y^{2}M_{\rm c}^{2}\frac{\Gamma[1+d/2]}{16\pi^{4+d/2}}\sum_{\mathbf{\bfell}}\frac{(1-(-1)^{2\mathbf{q}_{F}.\bfell})}{(\bfell.\bfell)^{1+d/2}},
\end{equation}
which converges. In 5 dimensions, for $d=1$, we find the result of
\cite{Antoniadis:1998sd}
\begin{equation}
\Delta M_{X}^{2}=y^{2}M_{\rm c}^{2}\frac{7\zeta(3)}{64\pi^{4}}.
\end{equation}
For $4+d$ dimensions, assuming that $q_{F}=1/2$ for all compact
dimensions we have 
\begin{equation}
\Delta M_{X}^{2}=y^{2}M_{\rm c}^{2}\frac{\Gamma[1+d/2]}{8\pi^{4+d/2}}\xi(1+d/2),
\end{equation}
where\footnote{The first few values of the multidimensional sums $\xi(s)$  are $\xi(3/2)=7\zeta(3)/4=2.1$ , $\xi(2)=6.6$,
$\xi(5/2)=14.0$, $\xi(3)=24.4$.} $\xi(s)=\sum_{\mathbf{\bfell}={\rm odd}}(\bfell.\bfell)^{-s}$.

 
\bigskip

\bibliographystyle{h-physrev5}

\end{document}